\def\fr#1#2{\hbox{${#1\over #2}$}}

\def\ni{\noindent}
\def\vs{\vskip.3cm}

\def\+{{(+)}}  \def\-{ {(-)} }   \def\0{ {(0)} }
\def\1{ {(1)} }  \def\2{ {(2)} }

\def\sq{Q\kern-6pt/}
\def\sQ{Q\kern-12pt\nearrow}
\documentclass[11pt,eqs]{article}

\usepackage{latexsym}

\def\be{\begin{equation}}             \def\ee{\end{equation}}
\def\ba{\begin{array}{rcl}}           \def\ea{\end{array}}
\def\beqa{\begin{eqnarray} }          \def\eeqa{\end{eqnarray} }
\def\beqalign{\begin{eqalign}}        \def\eeqalign{\end{eqalign}}
\def\leq#1{\label{eq:#1}}             \def\eq#1{(\ref{eq:#1})}
\def\bsubeq{\begin{subequations}}     \def\esubeq{\end{subequations}}
\def\bitem{\begin{itemize}}           \def\eitem{\end{itemize}}

\def\DJ{\leavevmode\setbox0=\hbox{D}\kern0pt
 \rlap{\kern.04em\raise.188\ht0\hbox{-}}D}
\def\dj{\leavevmode\setbox0=\hbox{d}\kern0pt
 \rlap{\kern.215em\raise.46\ht0\hbox{-}}d}

\newcommand{\bd}{\begin{displaymath}}
\newcommand{\ed}{\end{displaymath}}
\begin{document}

\title{Dilaton field induces commutative Dp-brane coordinate
\thanks{Work supported in part by the Serbian Ministry of Science and
Environmental Protection, under contract No. 1486.}}
\author{B. Sazdovi\'c\thanks{e-mail address: sazdovic@phy.bg.ac.yu}\\
       {\it Institute of Physics, 11001 Belgrade, P.O.Box 57, Serbia}}
\maketitle
\begin{abstract}
It is well known that space-time coordinates and corresponding Dp-brane
world-volume become non-commutative, if open string ends on Dp-brane with
Neveu-Schwarz background field $B_{\mu \nu}$. In this paper we extend these
considerations including the dilaton field $\Phi$, linear in coordinates
$x^\mu$. In that case the conformal part of the world-sheet metric appears as
new non-commutative variable and the coordinate in direction orthogonal to the
hyper plane $\Phi = const$, becomes commutative.
\end{abstract}
\vs

\ni {\it PACS number(s)\/}: 11.25.-w, 04.20.Fy, 11.10.Nx \par

\section{Introduction}

In the presence of the antisymmetric tensor field $B_{\mu \nu}$, quantization
of the open string ending on Dp-branes leads to non-commutativity of Dp-brane
world-volume. This result has been obtained for constant metric $G_{\mu \nu}$
and antisymmetric tensor $B_{\mu \nu}$, by some different methods: in terms of
mode expansion of the classical solution \cite{AC}, using conformal field
theory \cite{SW} and with the help of Dirac quantization procedure for
constraint systems \cite{ACL}.

In this paper we preserve the condition for background fields $G_{\mu \nu}$
and $B_{\mu \nu}$ to be constant, but we include linear part of the dilaton
field $\Phi$. This choice is consistent with the space-time field equation,
obtained from conformal invariance of the world-sheet theory.

In our choice of background, conformal
part of the world-sheet metric, $F$, is a dynamical variable. So, beside the
known boundary conditon $\gamma^{(0)}_i|_{\partial \Sigma}=0$, corresponding
to Dp-brane coordinate $x^i$ there is additional one $\gamma^{(0)}|_{\partial
\Sigma}=0$, corresponding to variable, $F$.

The noncommutative properties of the same theory, have been studied in
ref.\cite{BKM}. The authors used the mode expansion approach of ref.\cite{AC}.
They fixed, the conformal part of the metric, $F$, which we considered as a
variable of the theory. Consequently, they missed an additional boundary
condition, $\gamma^{(0)}|_{\partial \Sigma}=0$ and, as we will see later, loosed
generality of consideration.

In this paper, we apply canonical method and folloing ref.\cite{ACL}, we treat boundary
conditions as canonical constraints.  Using Dirac method, we find that consistency conditions lead to
two infinite sets of new constraints $\gamma^{(n)}_i|_{\partial \Sigma}=0$ and
$\gamma^{(n)}|_{\partial \Sigma}=0 \, , (n \geq 1)$. On one string endpoint we
substitute each set with one parameter conditions, $\Gamma_i (\sigma)=0$ and
$\Gamma (\sigma)=0$,  and on the other string endpoint with conditions ${\bar
\Gamma}_i (\sigma)=0$ and ${\bar \Gamma} (\sigma)=0$.

The periodicity condition solves the bar constraints. As a consequence of the
constraints $\Gamma_i (\sigma)=0$ and $\Gamma (\sigma)=0$, which are
particular orbifold conditions, all effective variables are symmetric under
transformation $\sigma \to - \sigma$, which reduces the phase space by half.

The constraints are of the second class. Instead to use Dirac brackets, as in
ref.\cite{ACL}, we explicitly solved the constraints in terms of effective
open string coordinate $q^i$ and effective open string conformal part of
the world-sheet metric  $f$.

We find effective theory in terms of open string variables. It has exactly the
same form as original theory, but with different Dp-brane background fields.
The explicit dependence on antisymmetric field disappears and it contributes
only to the effective metric tensor ${\tilde G}_{i j}$, as well as in the
absence of dilaton field. The effective dilaton field is linear in open string
coordinate, $q^i$.

We calculate Poisson brackets between all variables. We find that on the world-sheet boundary the
conformal part of the metric, $F$, does not commute with Dp-brane coordinates.
On the other hand, there exists one Dp-brane coordinate, $x \equiv a_\mu
x^\mu$, which commute with all other coordinates and with the world-sheet
metric, $F$.

\section{Definition of the model and canonical analysis}
\setcounter{equation}{0}

Let us start with the action of the bosonic open string,  \cite{FT}-\cite{P}
\be
S= \kappa  \int_\Sigma d^2 \xi \sqrt{-g} \left\{ \left[ {1 \over 2}g^{\alpha\beta}G_{\mu\nu}(x)
+{\varepsilon^{\alpha\beta} \over \sqrt{-g}}  B_{\mu\nu}(x)\right]
\partial_\alpha x^\mu \partial_\beta x^\nu + \Phi(x) R^{(2)} \right\} +
2\kappa \int_{\partial \Sigma} A_i d x^i   \,  ,   \leq{ac}
\ee
propagating in the non-trivial background, described by $x^\mu$ dependent
fields: metric $G_{\mu \nu}$, antisymmetric tensor field $B_{\mu\nu}=-
B_{\nu\mu}$, dilaton field $\Phi$ and $U(1)$ gauge field $A_i$ living on
Dp-brane. Here, $\xi^\alpha \, (\alpha=0,1)$ are coordinates of two
dimensional world-sheet $\Sigma$ and $x^\mu(\xi) \,  (\mu =0,1,...,D-1)$ are
coordinates of the D dimensional space-time $M_D$. We chose the gauge, where
$x^i \, (i=0,1,...p)$ are Dp-bane coordinates. The intrinsic world-sheet
metric we denote by $g_{\alpha \beta}$ and corresponding scalar curvature by
$R^{(2)}$. Through the paper we will use the notation $\partial_\alpha \equiv
{\partial \over \partial \xi^\alpha}$, $\partial_\mu \equiv {\partial \over
\partial x^\mu}$ and $\partial_i \equiv {\partial \over \partial x^i}$.

If both ends of the open string are attached to the same Dp-brane the action
can be written as
\be
S= \kappa  \int_\Sigma d^2 \xi \sqrt{-g} \left\{ \left[ {1 \over 2}g^{\alpha\beta}G_{\mu\nu}(x)
+{\varepsilon^{\alpha\beta} \over \sqrt{-g}}  {\cal F}_{\mu\nu}(x)\right]
\partial_\alpha x^\mu \partial_\beta x^\nu + \Phi(x) R^{(2)} \right\}
  \, ,   \leq{ac1}
\ee
where modified Born-Infeld field strength
\be
{\cal F}_{\mu\nu} = B_{\mu\nu} + (\partial_i A_j - \partial_j A_i) \delta_\mu^i
\delta_\nu^j   \,  ,
\ee
incorporate antisymmetric field with the field strength of the vector field.

In the conformal gauge
\be
g_{\alpha \beta} = e^{2 F} \eta_{\alpha \beta} \,  ,
\ee
we have $R^{(2)} = 2 \Delta F$, and the action takes the form
\be
S= \kappa  \int_\Sigma d^2 \xi  \left\{ \left[ {1 \over 2} \eta^{\alpha\beta} G_{\mu\nu}(x)
+ \varepsilon^{\alpha \beta}  {\cal F}_{\mu \nu}(x) \right]
\partial_\alpha x^\mu \partial_\beta x^\nu + 2 \Phi(x) e^{2 F}  \Delta F  \right\}
  \, .   \leq{ac2}
\ee
Note that the dilaton field breaks the conformal invariance, so that the
component $F$ of the metric tensor survives and the variables of the theory
are $x^\mu$ and $F$.

It is enormous task to make further progress with arbitrary background fields.
So, we are going to chose some particular solution of the space-time field
equations \cite{FCB}
\be
\beta^G_{\mu \nu} \equiv  R_{\mu \nu} - \fr{1}{4} {\cal F}_{\mu \rho \sigma}
{\cal F}_{\mu}{}^{\rho \sigma} +2 D_\mu a_\nu =0  \, ,
\ee
\be
\beta^{\cal F}_{\mu \nu} \equiv  D_\rho {\cal F}^\rho_{\mu \nu} -2 a_\rho
 {\cal F}^\rho_{\mu \nu} = 0  \, ,
\ee
\be
\beta^\Phi \equiv 4 \pi \kappa {D-26 \over 3} -
 R + \fr{1}{12} {\cal F}_{\mu \rho \sigma}
{\cal F}^{\mu \rho \sigma} - 4 D_\mu a^\mu + 4 a^2 = 0   \, ,
\ee
which are consequences of the world-sheet conformal invariance, as  necessary
conditions for consistency of the theory. Here, $a_\mu = \partial_\mu \Phi$,
${\cal F}_{\mu \rho \sigma}$ is field strength of the field ${\cal F}_{\mu
\nu}$ and $R_{\mu \nu}$, $R$ and $D_\mu$ are Ricci tensor, scalar curvature
and covariant derivative with respect to space-time metric. Following ref.
\cite{P} we chose
\be
G_{\mu \nu}(x) = G_{\mu \nu}=const \,  , \quad
{\cal F}_{\mu \nu} (x) = {\cal F}_{\mu \nu} = const \,   , \quad
\Phi(x) =\Phi_0 + a_\mu x^\mu  \,  , \quad (a_\mu = const)
\ee
which is exact solution for
\be
a^2 = \kappa \pi {26 - D \over 3}   \,  .
\ee

For simplicity, we suppose that antisymmetric tensor and gradient of dilaton
field are nontrivial only along directions of the Dp-brane world-volume, so
that  ${\cal F}_{\mu \nu} \to {\cal F}_{i j} $ and  $a_\mu \to a_i$. We also
chose coordinates so that  $G_{\mu \nu} = 0$ for $\mu =i \in \{0,1,..,p \}$
and $\nu =a \in \{p+1,...,D-1 \}$. So, the action under investigation is
\be
S= \kappa  \int_\Sigma d^2 \xi  \left\{  {1 \over 2} \eta^{\alpha\beta}
G_{\mu\nu} \partial_\alpha x^\mu \partial_\beta x^\nu
+ \varepsilon^{\alpha \beta}  {\cal F}_{i j}
\partial_\alpha x^i \partial_\beta x^j + 2 (\Phi_0 +a_i x^i) e^{2 F}  \Delta F  \right\}
  \,  ,   \leq{ac3}
\ee
and the components $x^a$, decouple from all other variables.

We are going to apply canonical methods to the action \eq{ac3}. Let us briefly
review the results of the canonical analysis, ref. \cite{BS1}, adapted to the
present case. The currents on the Dp-brane have the form
\be
J_\pm^i =P^T{}^{i j} j_{\pm j} +{a^i \over 2
a^2}i_\pm^\Phi = j_\pm^i -{ a^i \over a^2} j \,  ,
\qquad  (a_i \equiv \partial_i \Phi)                 \leq{J}
\ee
\be
i_\pm^F= {a^i \over a^2} j_{\pm i}-{1 \over 2 a^2} i_\pm^\Phi
\pm 2 \kappa {F^\prime } \,  , \qquad
i_\pm^\Phi= \pi \pm 2\kappa a_i x^{i \prime} \,  ,   \leq{FPh}
\ee
\be
j_{\pm i} =\pi_i +2\kappa \Pi_{\pm i j} {x^j}' \,  , \qquad
\left( \Pi_{\pm i j} \equiv  {\cal F}_{i j}  \pm {1 \over 2} G_{i j} \right)  \, \leq{jmi}
\ee
\be
j=a^i j_{\pm i} -{1 \over 2} i_\pm^\Phi =a^2 (i_\pm^F \mp 2 \kappa
F^\prime)  \,  ,    \leq{j}
\ee
where $\pi_i$ and $\pi$ are canonical momenta for $x^i$ and $F$.

For the directions orthogonal to the Dp-brane, only non trivial current is
\be
j_{\pm a} =\pi_a \pm \kappa G_{a b} {x^b}' \,   ,
\ee
where $\pi_a$ is momentum for $x^a$. It commutes with all other currents and
we will omit it from now.

We also introduce the projection operators
\be
P^L_{ i j} = {a_i a_j \over a^2} \,  ,
\qquad   P^T_{i j} = G_{i j} - {a_i a_j \over a^2}     \,   .
\leq{po}
\ee

All $\tau$ and $\sigma$ derivatives of the fields $x^i$ and $F$, can be
expressed in terms of the corresponding currents
\be
{\dot x}^i = {1 \over 2 \kappa} (J^i_- + J^i_+) \,  ,  \qquad
{\dot F}= {1 \over 4 \kappa} (i_-^F + i_+^F)   \, ,         \leq{td}
\ee
and
\be
x^{i \prime}= {1 \over 2 \kappa} (J^i_+ - J^i_- )  \, , \qquad
{F^\prime }= {1 \over 4 \kappa} (i_+^F - i_-^F)    \,   .   \leq{sd}
\ee

The canonical Hamiltonian density, which corresponds to Dp-brane part
\be
 {\cal H}_c =T_- -  T_+  \,   ,   \leq{hc}
\ee
is defined in terms of energy momentum tensor components
\be
T_\pm =\mp {1 \over 4\kappa} \left(G^{i j} J_{\pm i} J_{\pm j} + {j \over a^2}
 i_\pm^\Phi \right) +{1 \over 2} ( i_\pm^{\Phi \prime} - F^\prime i_\pm^\Phi)
  \,   .  \leq{emt}
\ee

The same chirality energy-momentum tensor components, satisfy two independent
copies of Virasoro algebras, while the opposite chirality components commute
\be
\{ T_\pm , T_\pm \}= -[ T_\pm(\sigma) +T_\pm({\bar \sigma}) ] { \delta^\prime} \,   ,
\qquad    \{ T_\pm , T_\mp \}= 0  \,  .     \leq{Vir}
\ee

\section{Open string boundary conditions as constraints}
\setcounter{equation}{0}

To describe open string evolution, both the equations of motion and the
boundary conditions are necessary. In paper \cite{BS1}, using canonical
method, we derived the world-sheet field equations, which in particular case
have the form, $\Delta x^\mu -2 a^\mu \Delta F = 0$ and $ a_\mu \Delta x^\mu
=0$. For $a^2 \neq 0$ they turn to the standard ones
\be
\Delta x^\mu = 0  \,  ,  \qquad  \Delta F = 0  \,  .
\ee

Let us consider the boundary conditions. It is useful to introduce the
variables
\be
\gamma_i^{(0)} \equiv {\delta S \over \delta x^{\prime i}} = \kappa (-G_{i
j} x^{j \prime}  +2 {\cal F}_{i j} {\dot x}^j -2 a_i
F^{\prime}) \,   ,   \qquad
\gamma^{(0)} \equiv {\delta S \over \delta F^{\prime}} =
 -2 \kappa a_i x^{i \prime }  \,  .   \leq{obc}
\ee
For the coordinates along Dp-brane directions and conformal part of the
world-sheet metric we use Neumann boundary conditions, allowing arbitrary
variations $\delta x^i$ and $\delta F$ on the string end points. Then the
boundary conditions can be written in the form
\be
\gamma_i^{(0)} |_{\partial \Sigma } =0 \,   ,  \qquad
 \gamma^{(0)} |_{\partial \Sigma } =0 \,    . \leq{cbc}
\ee
Comparing with dilaton free case, the second condition, relating to the
additional variable $F$, is a new one. Note that the constant field ${\cal
F}_{i j}$ does not appear in equations of motion and contributes only to the
boundary conditions.

For other coordinates we use Dirichlet boundary conditions, requiring the
edges of the string to be fixed, $\delta x^a |_{\partial \Sigma } =0$.

We are interested in the conditions  \eq{cbc}.  Using the expressions for
$\tau$ and $\sigma$ derivatives, \eq{td} and \eq{sd}, we can rewrite boundary
conditions in terms of the currents
\be
\gamma_i^{(0)} = \Pi_{+ i j} J_-^j + \Pi_{- i j} J_+^j + {a_i \over 2}
(i_-^F - i_+^F)  \,   , \qquad
\gamma^{(0)} = {1 \over 2} (i_-^\Phi - i_+^\Phi)  \,   . \leq{cobc}
\ee
Following approach of ref.\cite{ACL}, we will consider expressions,
$\gamma_i^{(0)} |_{\partial \Sigma }$ and $\gamma^{(0)} |_{\partial \Sigma}$,
as canonical constraints and we are going to find corresponding consistency
conditions. The fact that background fields $G_{i j}, {\cal F}_{i j}$ and $a_i
$ are $x^i$ independent, simplify Poisson brackets and we obtain
\be
\{ H_c, J_{\pm i} \} = \mp J_{\pm i}^\prime \,  , \qquad
\{ H_c, i^\Phi_\pm \} = \mp i^{\prime \Phi}_\pm \,  ,  \qquad
\{ H_c, i^F_\pm \} = \mp i^{\prime F}_\pm \,  .   \leq{cch}
\ee
Diarc consistency conditions generate two infinity sets of new conditions in
the form $\gamma_i^{(n)} |_{\partial \Sigma } =0 $ and $\gamma^{(n)}
|_{\partial \Sigma } =0$, $\,\,(n \geq 1)$, where
\be
\gamma_i^{(n)} \equiv \{ H_c, \gamma_i^{(n-1)} \}   = \partial^n_\sigma
\left\{\Pi_{+ i j} J_-^j + (-1)^n \Pi_{- i j} J_+^j +{a_i \over
2} \left[ i_-^F -(-1)^n i_+^F \right] \right\}  \,   ,
\ee
\be
\gamma^{(n)} \equiv \{ H_c, \gamma^{(n-1)} \}  = {1 \over 2} \partial^n_\sigma
\left[i_-^\Phi -(-1)^n i_+^\Phi \right] \, .
\ee
Using Taylor expansion, we can trade infinity derivatives at the point with
one variable function, so that conditions on one string endpoint take the form
\be
\Gamma_i(\sigma) \equiv \sum_{n \geq 0} {\sigma^n \over n!}
\gamma_i^{(n)}(0) = \Pi_{+ i j} J_-^j(\sigma)+ \Pi_{- i
j} J_+^j(-\sigma) +{a_i \over 2} \left[i_-^F(\sigma) -
i_+^F(-\sigma) \right]  \,  ,  \leq{v1}
\ee
\be
\Gamma (\sigma) \equiv \sum_{n \geq 0} {\sigma^n \over n!} \gamma^{(n)}(0) = {1 \over 2}
\left[i_-^\Phi(\sigma) - i_+^\Phi(-\sigma) \right]  \, .   \leq{v2}
\ee

On the other string endpoint, similarly we have
\be
{\bar \Gamma}_i(\sigma) \equiv \sum_{n \geq 0} {(\sigma - \pi)^n \over n!}
\gamma_i^{(n)}(\pi) = \Pi_{+ i j} J_-^j(\sigma)+ \Pi_{- i
j} J_+^j(2 \pi -\sigma) +{a_i \over 2} \left[i_-^F(\sigma) -
i_+^F(2 \pi -\sigma) \right]  \,  ,  \leq{v3}
\ee
\be
{\bar \Gamma} (\sigma) \equiv \sum_{n \geq 0} {(\sigma - \pi)^n \over n!}
\gamma^{(n)}(\pi) = {1 \over 2} \left[i_-^\Phi(\sigma) - i_+^\Phi(2 \pi -\sigma)
\right]  \, .   \leq{v4}
\ee
These expressions differ from boundary conditions, \eq{cobc}, only in the
arguments of positive chirality currents. On one string endpoint we have
$J_+^i (-\sigma) \, , \,\, i_+^F (-\sigma)$ and $i_+^\Phi (-\sigma)$, and on
the other one $J_+^i (2 \pi -\sigma) \, , \,\, i_+^F (2 \pi -\sigma)$ and
$i_+^\Phi (2 \pi -\sigma)$.

Because linear combinations of the constraints are also constraints, from
\eq{v1}-\eq{v4} we can conclude that all positive chirality currents are
periodic, for $\sigma \rightarrow \sigma + 2\pi$. Consequently, all variables
$x^i, \pi_i, F$ and $\pi$ are also periodic.

From \eq{cch} follows
\be
\{ H_c, \Gamma_{i}(\sigma) \} = \Gamma_{i}^\prime (\sigma) \,  , \qquad
\{ H_c, \Gamma (\sigma) \} = \Gamma^\prime (\sigma)  \,  ,  \leq{cchg}
\ee
which means that all constraints weakly commute with hamiltonian. Therefore,
there are no more constraints.

After same calculations we obtain
\be
\{\Gamma_{i}(\sigma) , \Gamma_{j}({\bar \sigma}) \} =
-\kappa \tilde{G}_{i j} \delta^\prime (\sigma - {\bar \sigma})  \,  ,  \qquad
\{\Gamma(\sigma) , \Gamma ({\bar \sigma}) \} = 0   \, ,  \leq{scc}
\ee
\be
\{\Gamma_{i}(\sigma) , \Gamma ({\bar \sigma}) \} =
- 2 \kappa a_i \delta^\prime (\sigma - {\bar \sigma})  \,  ,
\ee
were we introduced effective metric tensor
\be
{\tilde G}_{i j} \equiv  G_{i j} -4 {\cal F}_{i k}
P^{T k q} {\cal F}_{q j}   \,    .   \leq{efmt}
\ee
Following ref.\cite{SW} we will refer to it as the open string  metric tensor,
--- the  metric tensor seen by the open string.

The inverse of effective metric tensor we denote by ${\tilde G}^{i j}$. If we
raised the indices of covector $V_i$ with ${\tilde G}^{i j}$, we will put
tilde under corresponding vectors. So, we have $\tilde{V}^i = {\tilde G}^{i j}
V_j$, and $\tilde{V}^2 ={\tilde G}^{i j} V_i V_j$. We also preserve standard
notation, $V^i = G^{i j} V_j$ and $V^2 = G^{i j} V_i V_j$.

Direct calculation yields
\be
\{\Gamma_A (\sigma) , \Gamma_B ({\bar \sigma})\} = - \kappa
\left| \begin{array}{clcr}
 \tilde{G}_{i j} & 2 a_i  \\
2 a_j &   0
\end{array}  \right|
 \delta^\prime (\sigma - {\bar \sigma}) \equiv  \Delta_{A B}
 \delta^\prime (\sigma - {\bar \sigma}) \,  ,
\ee
and
\be
\triangle \equiv \det \triangle_{A B} = - 4 (-\kappa)^{p+2}
\tilde{a}^2 \det \tilde{G}_{i j}   \,   ,
\ee
where we denoted $\Gamma_A = \{ \Gamma_i \, , \Gamma \}$. Therefore, for
$\tilde{a}^2 \neq 0$, which we assume, we have $rank \triangle_{A B} = p+2$
and all constraints are of the second class (except the zero mode, see
\cite{BS3}).

\section{Solution of the boundary conditions}
\setcounter{equation}{0}

The periodicity condition solves the second set of constraints \eq{v3}-\eq{v4}
and we are going to solve the first one \eq{v1}-\eq{v2}. It is useful to
introduce new variables
\be
q^i (\sigma) = {1 \over 2} \left[x^i (\sigma) + x^i (-\sigma) \right] \,  ,
\quad  {\bar q}^i (\sigma) = {1 \over 2} \left[x^i (\sigma) -
 x^i (-\sigma) \right] \,  ,   \leq{nv1}
\ee
\be
p_i (\sigma) = {1 \over 2} \left[\pi_i (\sigma) + \pi_i (-\sigma) \right] \,  ,
\quad  {\bar p}_i (\sigma) = {1 \over 2} \left[\pi_i (\sigma) -
\pi_i (-\sigma) \right] \,  ,    \leq{nv2}
\ee
\be
f(\sigma) = {1 \over 2} \left[F (\sigma) + F (-\sigma) \right] \,  ,
\quad  {\bar f} (\sigma) = {1 \over 2} \left[F (\sigma) -
F(-\sigma) \right] \,  ,          \leq{nv3}
\ee
\be
p (\sigma) = {1 \over 2} \left[\pi (\sigma) + \pi (-\sigma) \right] \,  ,
\quad  {\bar p} (\sigma) = {1 \over 2} \left[\pi (\sigma) -
\pi (-\sigma) \right] \,  ,       \leq{nv4}
\ee
to which we will referee as open string variables.

Using the relations
\be
{1 \over 2} [ j_{- i}(\sigma) + j_{+ i}(-\sigma) ]= p_i +
2 \kappa {\cal F}_{i j} {\bar q}^{j \prime }  - \kappa G_{i j} q^{j \prime }  \,  ,
\ee
\be
{1 \over 2} [ j_{- i}(\sigma) - j_{+ i}(-\sigma)]= {\bar p}_i +
2 \kappa {\cal F}_{i j} q^{j \prime}  - \kappa G_{i j} {\bar q}^{ j \prime}  \,  ,
\ee
\be
{1 \over 2} [ i^\Phi_-(\sigma) - i^\Phi_+ (-\sigma)]= {\bar p}
- 2 \kappa a_i {\bar q}^{ i \prime}   \,  ,
\ee
we can write the constraints in terms of open string variables
\be
\Gamma_i (\sigma) = 2 ({\cal F} P^T )_i{}^j p_j + {\bar p}_i + {1
\over a^2} {\cal F}_{i j} a^j p - \kappa {\tilde G}_{i j} {\bar
q}^{ j \prime} -2 \kappa a_i {\bar f}^\prime  \,   ,
\ee
\be
\Gamma (\sigma) = {\bar p} -2 \kappa a_i {\bar q}^{i \prime }  \,   .
\ee

The symmetric and antisymmetric parts under transformation $\sigma \to
-\sigma$, separately vanish. Therefore, form $\Gamma_i (\sigma) = 0$, we
obtain
\be
{\bar p}_i =0 \,  , \qquad 2 ({\cal F} P^T )_i{}^j p_j  + {1 \over
a^2} {\cal F}_{i j} a^j p - \kappa {\tilde G}_{i j} {\bar
q}^{j \prime } -2 \kappa a_i {\bar f}^\prime =0  \,   ,   \leq{c1}
\ee
and from $\Gamma (\sigma) = 0$
\be
{\bar p} =0   \,   ,   \qquad     a_i {\bar q}^{i \prime } = 0   \,   .  \leq{c2}
\ee

We can solve all antisymmetric (bar) variables in terms of symmetric ones
\be
{\bar p}_i =0 \,  , \qquad  {\bar q}^{i \prime } = -2( \Theta^{i
j} p_j + \Theta^i p)   \,  ,  \leq{s1}
\ee
\be
{\bar p} =0 \,  , \qquad  {\bar f}^{\prime} = 2 \Theta^i p_i   \,  ,  \leq{s2}
\ee
where
\be
\Theta^{i j} = {-1 \over \kappa} \tilde{P}^T{}^{i k} {\cal F}_{k q} P^T{}^{q j}
\,  ,   \qquad (\Theta^{i j} = - \Theta^{j i} )  \leq{tmn}
\ee
\be
\Theta^i = {({\tilde a} {\cal F})^i \over 2 \kappa  {\tilde a}^2 } =
{( a {\cal F} {\tilde G}^{-1})^i \over 2 \kappa  a^2 }  \,  ,  \leq{ti}
\ee
and in analogy with \eq{po} we introduced tilde projectors
\be
\tilde{P}^{L i j} = {\tilde{a}^i \tilde{a}^j \over \tilde{a}^2} \,  ,
\qquad   \tilde{P}^{T i j} = {\tilde G}^{i j} -
{\tilde{a}^i \tilde{a}^j \over \tilde{a}^2}   \,   .
\ee

Using  \eq{nv1}- \eq{nv4} and  \eq{s1}-\eq{s2}, we can express original
variables in terms of new ones
\be
x^i = q^i -2 \int^\sigma d \sigma_1  \left( \Theta^{i
j} p_j + \Theta^i p \right)  \,    , \qquad \pi_i = p_i
\,  ,    \leq{sx}
\ee
\be
F= f + 2 \, \Theta^i \int^\sigma d
\sigma_1 \, \, p_i \,  ,    \qquad  \pi = p    \,  .    \leq{sf}
\ee

As a consequence of particular form of the conditions, $\Gamma_i (\sigma) =0$
and $\Gamma(\sigma)=0$, the effective theory depends only on the variables
symmetric under $\sigma \to - \sigma$.

\section{The effective theory in terms of open string variables}
\setcounter{equation}{0}

The original string theory is completely described by the energy-momentum
tensor $T_\pm$, \eq{emt}, in terms of variables $x^i, \, F$ and theirs momenta
$\pi_i, \, \pi$. We are going to find effective energy-momentum tensor
${\tilde T}_\pm$, in terms of new variables $q^i, \, f$ and corresponding
momenta $p_i, \, p$.

Because we have expression for energy-momentum tensor in terms of the
currents, we will first express the currents in terms of new variables. In
analogy with equations \eq{FPh}, \eq{jmi}, \eq{j} and  \eq{J} we introduce
new, open string currents
\be
\tilde{i}_\pm^\Phi= p \pm 2 \kappa a_i q^{i \prime } \,  ,   \leq{tFPh}
\ee
\be
\tilde{j}_{\pm i} =p_i \pm  \kappa \tilde{G}_{i j} q^{j \prime }    \,  ,
\qquad
\tilde{j}= \tilde{a}^i \tilde{j}_{\pm i} -{1 \over 2} \tilde{i}_\pm^\Phi
 \,   ,   \leq{tj}
\ee
and
\be
\tilde{J}_{\pm}^{i} = \tilde{P}^T{}^{i j} \tilde{j}_{\pm j}
+{\tilde{a}^i \over 2 \tilde{a}^2} \tilde{i}_\pm^\Phi = \tilde{G}^{i j}
\tilde{j}_{\pm i} -{ \tilde{a}^i \over \tilde{a}^2} \tilde{j} \,  .
\leq{tJ}
\ee
They depend on new variables in similar way as the original currents depend on
the original variables. The metric tensor is substituted by the effective one
and the main difference is, that there is no explicit dependence on
antisymmetric tensor, but it contributes only through the effective metric
tensor. Formally, we can first put ${\cal F}_{i j} \to 0$ and then $G_{i j}
\to \tilde{G}_{i j}$. Let us stress that in all open string currents we
systematically substitute $a^i$ and $a^2$ with $\tilde{a}^i$ and $
\tilde{a}^2$.

With the help of  \eq{sx}-\eq{sf}, we can express the original currents in
terms of new variable. We will preserve the same notations for these
expressions and relate them with open string currents \eq{tFPh}-\eq{tJ},
obtaining
\be
i_\pm^\Phi =  \tilde{i}_\pm^\Phi  \,  , \qquad
{j \over a^2} = { \tilde{j} \over \tilde{a}^2} + {2 \kappa \over a^2}
a^i {\cal F}_{i j}  q^{j \prime } \,  ,   \leq{cip}
\ee
\be
J_{\pm i} = \pm 2 {\tilde \Pi}_{\pm i j} \tilde{J}_\pm^j   \,   .
\qquad   \left( {\tilde \Pi}_{\pm i j} =  \Pi_{\pm i j} -
P^L{}_i{}^k {\cal F}_{k j} \right)    \leq{cJ}
\ee

Using the expression
\be
4 G^{i j} {\tilde \Pi}_{\pm i k} {\tilde \Pi}_{\pm j
q}= \tilde{G}_{k q} \pm 2 ({\cal F}_{k r}
P^{L r}{}_q - P^L{}_k{}^r  {\cal F}_{r q})  \,    ,
\ee
we get useful relation
\be
G^{i j} J_{\pm i} J_{\pm j} =  \tilde{G}_{i j}
\tilde{J} _\pm^i \tilde{J} _\pm^j  \mp {2 \over \tilde{a}^2 }
\tilde{i}_\pm^\Phi \tilde{a}^i {\cal F}_{i j} \tilde{j}_\pm^j  \,
.   \leq{ur}
\ee

Finally, we are in position to find energy-momentum tensor in terms of open
string variables. With the help of  \eq{cip}-\eq{ur} we obtain
\be
T_\pm = \tilde{T}_\pm   \,  ,  \leq{ttt}
\ee
where
\be
\tilde{T}_\pm =\mp {1 \over 4\kappa} \left( \tilde{G}^{ij} \tilde{J}_{\pm
i} \tilde{J}_{\pm j} + {\tilde{j} \over \tilde{a}^2}  \tilde{i}_\pm^\Phi
\right) + {1 \over 2} ( \tilde{i}_\pm^{\Phi \prime } - f^\prime
\tilde{i}_\pm^\Phi)
  \,   .  \leq{nemt}
\ee

So, we can conclude that the effective energy-momentum tensor depend on open
string currents in exactly the same way as the original energy-momentum tensor
depend on original currents.

Therefore, instead of standard formulation in terms of variables  $x^i, F$ and
corresponding momenta $\pi_i , \pi$ in background fields $G_{i j}, \, {\cal
F}_{i j}$ and $\Phi$ the effective theory is expressed in term of new
variables $q^i, f$ and corresponding momenta $p_i, p$ in background fields
${\tilde G}_{i j}, {\tilde {\cal F}}_{i j}=0 $ and ${\tilde \Phi}=\Phi_0 + a_i
q^i$. In the first case, together with equations of motion, the boundary
conditions \eq{cbc} must be used. In the second case, we should impose the
symmetries $\sigma \to \sigma + 2 \pi$ and  $\sigma \to - \sigma$, which are
some particular forms of orbifold conditions.

The open string hamiltonian takes the form $\tilde{\cal H}_c = \tilde{T}_- -
\tilde{T}_+$, and corresponding equations of motion are
\be
\tilde{\Delta} q^i  =0  \,  , \qquad   \tilde{\Delta} f = 0   \,  .   \leq{olJ}
\ee
The Laplace operator  $\tilde{\Delta}$ is defined with field $f$ as a
conformal part of the effective world-sheet metric, ${\tilde g}_{\alpha \beta}
= e^{2 f} \eta_{\alpha\beta}$.

\section{Non-commutative world-sheet metric and commutative Dp-brane direction}
\setcounter{equation}{0}

From standard Poisson brackets
\be
\{ x^i (\sigma) , \pi_j ({\bar \sigma}) \} =
\delta^i _j \delta(\sigma- {\bar \sigma})  \,   ,  \qquad
\{ F (\sigma) , \pi ({\bar \sigma}) \} =
 \delta(\sigma- {\bar \sigma})  \,  ,
\ee
and  relations \eq{nv1}- \eq{nv4} we have
\be
\{ q^i (\sigma) , p_j ({\bar \sigma}) \} =
\delta^i _j \delta_s(\sigma , {\bar \sigma})   \,  ,  \qquad
\{ f (\sigma) , p ({\bar \sigma}) \} =
\delta_s(\sigma , {\bar \sigma})   \,  ,
\ee
where
\be
\delta_s(\sigma , {\bar \sigma}) = {1 \over 2} \left[\delta(\sigma- {\bar \sigma})
+ \delta(\sigma + {\bar \sigma}) \right]    \,  , \qquad  ( \sigma, {\bar \sigma}
\in [0,\pi])
\ee
is symmetric delta-function. So, $q^i$ and $p_i$, as well as $f$ and $p$, are
canonically conjugate variables on subspace symmetric under $\sigma \to -
\sigma$.

With the help of \eq{sx}-\eq{sf} we can calculate Poisson brackets between
dynamical variables
\be
\{ x^i (\sigma) , x^j ({\bar \sigma}) \} = 2
\Theta^{i j} \Delta(\sigma + {\bar \sigma})  \,  ,  \qquad
\{ x^i (\sigma) , F ({\bar \sigma}) \} = 2
\Theta^i \Delta(\sigma + {\bar \sigma})  \,  ,
\ee
where $\Theta^{i j}$ and $\Theta^i$ have been defined in \eq{tmn} and \eq{ti}
respectively and
\be
\Delta (\sigma + {\bar \sigma}) = \theta(\sigma + {\bar \sigma})
= \left\{ \begin{array}{rrr}
0 & \mbox{$\sigma=0={\bar \sigma}$} \\
1 &  \mbox{$\sigma=\pi={\bar \sigma}$} \\
\fr{1}{2} &  \mbox{otherwise} \,\,\,
\end{array}
 \right.      \qquad  .
\ee

It is useful to separate the center of mass, $x^i_{cm} = {1 \over \pi}
\int_0^\pi d \sigma x^i (\sigma)$, in the form $x^i (\sigma) = x^i_{cm} + X^i
(\sigma)$, so that we have
\be
\{ X^i (\sigma) , X^j ({\bar \sigma}) \} = 2
\Theta^{i j} \left[ \Delta(\sigma + {\bar \sigma}) - {1 \over 2} \right]
= \Theta^{i j}
\left\{ \begin{array}{rrr}
-1 & \mbox{$\sigma=0={\bar \sigma}$} \\
1 &  \mbox{$\sigma=\pi={\bar \sigma}$} \\
0 &  \mbox{otherwise} \,\,\,
\end{array}
 \right.  \qquad  ,  \leq{xx}
\ee
\be
\{ X^i (\sigma) , F ({\bar \sigma}) \} = 2
\Theta^i \left[ \Delta(\sigma + {\bar \sigma}) - {1 \over 2} \right]
= \Theta^i \left\{ \begin{array}{rrr}
-1 & \mbox{$\sigma=0={\bar \sigma}$} \\
1 &  \mbox{$\sigma=\pi={\bar \sigma}$} \\
0 &  \mbox{otherwise} \,\,\,
\end{array}
 \right.  \qquad  .    \leq{xF}
\ee

The relation \eq{xF} has not be considered before in the
literature. It shows that in the presence of dilaton field, the
non-commutativity between coordinates and conformal part of the world-sheet
metric appears on the world-sheet boundary. The expression for this new
non-commutativity parameter, $\Theta^i$, is proportional to Born-Infeld field, ${\cal
F}_{i j}$.

The relation \eq{xx} has the same form as in the absence of dilaton field
\cite{AC}-\cite{BKM}, but there are some significant differences.

Let us first explain geometrical meaning of the projectors $P^T{}^{i j}$ and
$\tilde{P}^T{}^{i j}$. Note that vector $a_i$ is normal  to the $D-1$
dimensional submanifold $M_{D-1}$, defined by the condition $\Phi(x)= const$.
For $a^2 \neq 0$ $({\tilde a}^2 \neq 0)$, the corresponding unit vectors for
the closed and open string respectively are $n_i = {a_i \over
\sqrt{\varepsilon a^2}}$ and ${\tilde n}_i = {a_i \over
\sqrt{\tilde{\varepsilon} {\tilde a}^2}}$.  Here $\varepsilon=1 \,
(\tilde{\varepsilon}=1)$ if $a_i$ is time like vector, and $\varepsilon=-1 \,
(\tilde{\varepsilon}=-1)$ if $a_i$ is space like vector with respect to
metrics $G_{i j} ( \tilde{G}_{i j} )$. Consequently,
\be
P^T{}_{i j} = G_{i j}- \varepsilon n_i n_j \equiv G_{i j}^{(D-1)} \,  , \qquad
{\tilde P}^T{}_{i j} = {\tilde G}_{i j}- \varepsilon {\tilde n}_i {\tilde n}_j
\equiv {\tilde G}_{i j}^{(D-1)} \,  ,  \leq{im}
\ee
are  induced metrics on $M_{D-1}$, and we can rewrite \eq{tmn} in the form
\be
\Theta^{i j} = {-1 \over \kappa} \tilde{G}^{(D-1)}{}^{i k}
{\cal F}_{k q} G^{(D-1)}{}^{q j}   \,  .    \leq{tmn1}
\ee
This expression in fact is similar as in the absence of dilaton field. The
essential part again is Born-Infeld field strength ${\cal F}_{k q}$, but in
the present case we raised indices  with metrics of submanifold $M_{D-1}$:
$\tilde{G}^{(D-1)}{}^{ij}$ and $G^{(D-1)}{}^{ij}$ instead of metrics of
manifold $M_D$: $G_{eff}^{ij} = (G -4 {\cal F} G^{-1} {\cal F})^{-1}{}^{ij} $
and $G^{ij}$.

From the relations $a_i P^{T i j} = 0$ and ${\tilde a} {\cal F} a =0$, it
follow $a_i \Theta^{i j} = 0$ and $a_i \Theta^i = 0$, so that the component $x
\equiv a_i x^i$ commutes with all other coordinates as well as with the
conformal part of the metric
\be
\{ x (\sigma) , x^j ({\bar \sigma}) \} = 0    \,  , \qquad \{ x (\sigma) , F ({\bar \sigma}) \} = 0    \,  .
\ee
This is an example of Dp-brane with one commutative coordinate in $a_i$
direction (proportional to gradient of the dilaton field).

\section{Concluding remarks}
\setcounter{equation}{0}

In this paper we presented an interesting example, choosing background with
dilaton field linear in coordinate and constant metric and antisymmetric
fields. This chose of background preserves the conformal symmetry of the
world-sheet theory. We investigated the contribution of dilaton field to the
non-commutativity of the Dp-brane world-volume.

We found complete set of constraints $\Gamma_j(\sigma) =0$, $\Gamma(\sigma)
=0$, ${\bar \Gamma}_j(\sigma) =0$ and ${\bar \Gamma}(\sigma) =0$ which include
open string boundary conditions as canonical constraints, and infinite sets of
constraints obtained from Dirac consistency conditions. We solved them
explicitly, imposing periodically condition  and expressing all variables odd
under $\sigma \to - \sigma$ in terms of the even ones.

The effective theory in terms of open string variables $q^j$ and $f $, has
precisely the same form as the original theory in terms of closed string
variables $x^j$ and $F$. It has the same form of energy-momentum tensor, the
same form of hamiltonian and the same form of field equations. There are two
differences. First, the closed string background $G_{i j}, \, {\cal F}_{i j} =
B_{i j} + \partial_i A_j - \partial_j A_i $ and $\Phi = \Phi_0 + a_i x^i$
should be substituted by the open string one
\be
G_{i j} \to {\tilde G}_{i j} =  G_{i j} -4 {\cal F}_{i k}
P^{T k q} {\cal F}_{q j}   \,   , \qquad
{\cal F}_{i j}  \to \tilde{\cal F}_{i j} = 0
\,  ,   \qquad \Phi  \to {\tilde \Phi}= \Phi_0 + a_i q^i  \, .   \leq{csb}
\ee
Second, for open string variables $q^i$ and $f$, instead of the boundary
conditions $\gamma^{(0)}_i|_{\partial \Sigma}=0$ and $\gamma^{(0)}|_{\partial
\Sigma}=0$, the symmetries under $\sigma \to \sigma + 2\pi$ and $\sigma \to -
\sigma$ should be imposed.

The relation between the closed and open string variables clarify the origin
of non-commutativity. The closed string variables depend on open string
variables, but also on the corresponding momenta. So, the Poisson brackets
between the variables are nontrivial on the world-sheet boundary.

Beside known coordinate non-commutativity, we established the
non-commutativity relation between coordinates and conformal part of the
world-sheet metric. We obtained explicit expressions for non-commutativity
parameters $\Theta^{i j}$ \eq{tmn}, and $\Theta^i$ \eq{ti}, and found that
both are proportional to the Born-Infeld field strength ${\cal F}_{i j}$. In the
presence of dilaton field linear in coordinates we have, $a_i \Theta^{i j}=0$
and $a_i \Theta^i=0$. Therefore, it turns one coordinate, in $a_i$ direction,
to the commutative one.

Let us compare the results of the present paper with that of ref.\cite{BKM},
where the same action has been investigated. First, they fixed the conformal part
of the metric, $F$, which in our case is variable of the theory. Second, from conformal invariance
condition on the boundary, $(T_+ + T_-) |_{\partial \Sigma}=0$,
they obtained additional constraint on background fields $a^i {\cal F}_{i j} =
0$.

In our approach, the boundary condition $(T_+ + T_-) |_{\partial \Sigma}=0$ is satisfy
for arbitrary background fields. In fact with the help of \eq{ttt}, we have $T_\pm = \tilde{T}_\pm $, and
above equation takes the form $ {\tilde T}_+ + {\tilde T}_- =0$. As a
consequence of the second relation \eq{csb} this condition is satisfy without
any restriction on the background fields. Therefore, boundary conditions which
we obtained  from the action principle, are enough to fulfill requirement
that there is no net flow of energy and momentum from the boundary.

The constraint of ref.\cite{BKM} on background fields, $a^i {\cal F}_{i j} =
0$, in fact is consequence of their gauge fixing, $F=0$.
In this gauge the Poisson bracket between
$x^i$ and $F$ must vanish, which according to the relation \eq{ti}
produces the above constraint.

In the present paper, the effective metric tensor, ${\tilde G}_{i j}$, and non-commutativity
parameters, $\Theta^{i j}$ and $\Theta^i$, explicitly depend on the dilaton
field and the existence of commutative coordinate appears for arbitrary
background fields. In ref.\cite{BKM} the commutative direction is consequence
of the relation, $a^i {\cal F}_{i j} = 0$. Particularly, on this relation our
effective metric tensor and non-commutativity parameter $\Theta^{i j}$ turn to
ones of ref.\cite{BKM}, while the parameter $\Theta^i$ vanishes. So, our
results are more general, because they valid without above restriction on
background fields.

It is instructive to compare the symmetric and antisymmetric string
parameters, in three different cases. For closed string, the metric tensor and
Born-Infeld field strength satisfies
\be
{\cal F}^{i j} \pm \fr{1}{2} G^{i j} = (G^{-1} \Pi_\pm G^{-1})^{i j}  \,  .
\ee
The open string is sensitive to the effective metric tensor and
non-commutative parameter. In dilaton free case they produce
\be
\kappa \theta^{i j} \pm \fr{1}{2} G_{eff}^{i j} = (G^{-1} \Pi_\pm G^{-1}_{eff})^{i j}
\, ,
\ee
while in the case of the present paper, corresponding relation obtains the
form
\be
\kappa \Theta^{i j} \pm \fr{1}{2} {\tilde G}_{D-1}^{i j} =
(G_{D-1}^{-1} \Pi_\pm {\tilde G}_{D-1}^{-1})^{i j}  \,  .
\ee
Here, $G^{eff}_{ij} = (G -4 {\cal F} G^{-1} {\cal F})_{ij} $ and ${\tilde
G}_{i j} =  (G -4 {\cal F} G_{D-1}^{-1} {\cal F})_{i j}$ are effective metric
tensors in absence and presence of dilaton field, respectively. Therefore, the
addition of dilaton field just turns metric $G_{i j}$ of $M_D$ to the metric
$G^{D-1}_{i j}$ of $M_{D-1}$.

\end{document}